\documentclass[oneside,english]{amsart}
\usepackage[T1]{fontenc}
\usepackage[latin9]{inputenc}
\usepackage{geometry}
\usepackage{color}
\usepackage{amsthm}
\usepackage{amssymb}
\usepackage{stmaryrd}

\makeatletter

\newcommand*\LyXZeroWidthSpace{\hspace{0pt}}

\numberwithin{equation}{section}
\theoremstyle{plain}
\newtheorem{thm}{\protect\theoremname}[section]
\theoremstyle{remark}
\newtheorem{rem}[thm]{\protect\remarkname}
\newenvironment{lyxlist}[1]
	{\begin{list}{}
		{\settowidth{\labelwidth}{#1}
		 \setlength{\leftmargin}{\labelwidth}
		 \addtolength{\leftmargin}{\labelsep}
		 }}
	{\end{list}}
\theoremstyle{remark}
\newtheorem*{acknowledgement*}{\protect\acknowledgementname}

\makeatother

\usepackage{babel}
\providecommand{\acknowledgementname}{Acknowledgement}
\providecommand{\remarkname}{Remark}
\providecommand{\theoremname}{Theorem}

\begin{document}
\title[Signature Change Through a Natural Extension of Kaluza-Klein Theory]{A Conceptual Introduction To Signature Change Through a Natural Extension
of Kaluza-Klein Theory}
\author{Vincent Moncrief and Nathalie E. Rieger}
\address{Department of Mathematics and Physics, Yale University}
\email{vincent.moncrief@yale.edu}
\address{Department of Mathematics, Yale University}
\email{n.rieger@yale.edu}
\begin{abstract}
We propose an extension of basic Kaluza-Klein theory in which the
higher-dimensional Lorentzian manifold develops a Cauchy horizon rather
than remaining globally hyperbolic as in the conventional framework.
In this setting, the $U(1)$-generating Killing field, assumed to
exist in Kaluza-Klein theory, undergoes a transition in its causal
character, from spacelike in the globally hyperbolic region to timelike
in an acausal extension through a horizon. This yields a (lower-dimensional)
quotient manifold whose metric changes signature from Lorentzian to
Riemannian. In this way, one observes a singular, signature changing
transition emerging rather naturally from the projection of a globally
smooth, even analytic, Lorentzian geometry ``up in the bundle''.
This reveals a ``signature change without signature change'' scenario---a
phrasing inspired by John Wheeler---and extends the usual Kaluza-Klein
framework in a conceptually natural direction.
\end{abstract}

\maketitle

\section{Introduction}

An interesting extension of conventional semi-Riemannian geometry
allows, among other possibilities, for a symmetric, $2$nd rank tensor
field to undergo a singular transition from defining Lorentzian geometry
in some open region of a manifold $M$ to defining
Riemannian geometry in a complementary, open region with the transition,
on which the metric tensor must degenerate, occurring on an embedded
hypersurface $\mathcal{H}\subset M$~\cite{Dray, Hasse + Rieger-Transformation, Hayward, Kossowski - The Volume BlowUp and Characteristic Classes for Transverse, Kriele + Martin - Black Holes ..., Rieger+Hasse-Loops}.
While this phenomenon can be analyzed abstractly, for purely differential
geometric reasons~\cite{Larsen, Rieger}, or within the framework of mathematical physics~\cite{Ellis - Change of signature in classical relativity, Hellaby et al, Martin, Martin Hamiltonian}, including Hamiltonian treatments of signature change in general relativity, we wish to point out that it can arise spontaneously
through a straightforward extension of the Kaluza-Klein generalization
of Einstein's general relativity theory~\cite{Ellis + Piotrkowska, Kerner}. In this scenario, the lower
dimensional quotient space undergoes a signature change of the type
mentioned above, while an actual metric on the total space, ``up
in the bundle'', remains Lorentzian throughout and satisfies the
Einstein field equations but transitions from being globally hyperbolic
to causality violating across a so-called Cauchy horizon. In this
setting, a preferred Killing field, assumed to exist by the Kaluza-Klein
formulation, transitions from being spacelike (in the globally hyperbolic
region) to timelike (in the acausal extension) while becoming null
and tangential to the Cauchy horizon's null generators at the interface.
Thus one achieves ``signature change without signature change''---a
phrasing inspired by John Wheeler---while extending the usual Kaluza-Klein
framework in a straightforward way.\footnote{Recall that John Wheeler often described primordial black holes as
exhibiting ``mass without mass'' or ``charge without charge'',
since they incorporated one or both of these qualities without actually
entailing material bodies having either mass or charge.}

~

Examples of this phenomenon already exist in the $4$-dimensional
context of Taub-NUT (Newman, Unti, Tamburino)-like spacetimes, which
contain compact Killing horizons of the aforementioned type whereby
the corresponding, $3$-dimensional quotient manifolds undergo the
signature change in question. For these cases, where the Einstein
field equations are being enforced up in the $4$-dimensional bundle,
there are rather surprising theorems that ensure that the existence
of a Cauchy horizon actually implies the presence of an associated
Killing symmetry of the transitioning type~\cite{3 Symmetries of cosmological Cauchy horizons with exceptional orbits,4 On spacetimes containing Killing vector fields with non-closed orbits,1 Symmetries of cosmological Cauchy horizons,2 Symmetries of Cosmological Cauchy Horizons with Non-Closed Orbits}. 

~

On the other hand, there are related constructions that show (at least
in the case of analytic metrics) that Einstein spacetimes that develop
such Cauchy horizons are highly non-generic, even within the context
of solutions to the field equations of the same isometry class. In
Hamiltonian language, these ``generalized Taub-NUT'' spacetimes
exhaust only a Lagrangian submanifold of the associated phase space
of solutions possessing the Killing symmetry imposed via the Kaluza-Klein
paradigm~\cite{6 Moncrief ,5 The space of generalized Taub-Nut spacetimes}.
A motivation for these earlier studies was to provide indirect support
for the cosmic censorship conjecture for Einstein's theory---often
regarded as the main open mathematical problem of general relativity.
The occurrence of such causality-violating extensions to globally
hyperbolic Einstein spacetimes would, if they proved to fill out an
open subset (in some suitable function space topology) of the space
of all solutions on a given manifold, could disprove the cosmic censorship
conjecture, at least for that manifold. But the fact that such generalized
Taub-NUT spacetimes necessarily admit Killing symmetries at all, and
indeed constitute only the aforementioned Lagrangian submanifold of
solutions in that isometry class, provides strong support for the
cosmic censorship idea. 

~

From the standpoint of providing examples of signature-changing geometries,
though, albeit only ones subject to the Einstein field equations,
their lack of genericity is perhaps only a peripheral issue. For applications
to potentially physically interesting spacetimes, Kaluza-Klein theory
normally posits a higher than four dimensional Lorentzian manifold,
with a metric subject to the Einstein field equations, and imposes
the existence of a (typically spacelike) isometry group thereon in such a way that the corresponding quotient $4$-manifold rather
miraculously satisfies a variant of either the Einstein-Maxwell-scalar
field equations (descending from a $5$-dimensional bundle and $U(1)$
isometry group) or even the Einstein-Yang-Mills wavemap field equations
when the initial manifold is of still higher dimension and a suitable,
non-abelian isometry group is imposed upon the metric.

~

In this conventional Kaluza-Klein scenario, the spacelike character
of the imposed isometry group ensures that the resulting quotient
$4$-manifold is uniformly Lorentzian and can thus provide a potential
model for a physical universe, at least at this classical level of
analysis. But, if instead, a Cauchy horizon develops up in the bundle,
and if the aforementioned theorems extend to apply in this higher-dimensional
setting then one would expect to see a corresponding signature change
down in the base. Fortunately, as we shall see, a number of the relevant
theorems do extend to these higher-dimensional settings. As far as
we know, however, those asserting the existence of (a Lagrangian submanifold
of) generalized Taub-NUT spacetimes have not yet been so extended,
though there is good reason to suppose that this can be done. For
analytic metrics the main tool used in $4$-dimensions was the Cauchy-Kowalewski
theorem which of course is applicable in any dimension. In any event,
there are explicitly known examples of higher-dimensional Einstein
spacetimes that develop compact Cauchy horizons across which the requisite
Killing field changes type from spacelike (in the globally hyperbolic
region) to null (on the horizon where it is tangent to the horizon's
null generators) to timelike (in the acausal extension, which admits
closed timelike curves). The simplest of these is the product of a
flat Riemannian torus $\{T^{n};e\}$ for $n\geq3$ with Misner's two-dimensional
model for Taub-NUT behavior defined on a Lorentzian cylinder diffeomorphic
to $S^{1}\times\mathbb{R}$. The resulting Einstein (in fact flat)
spacetime has dimension $n+2$ and admits the desired type-changing
Killing field and $(n+1)$-dimensional Cauchy horizon diffeomorphic
to $T^{n+1}$. One would expect this to be a very special case of
$(n+2)$-dimensional generalized Taub-NUT solutions definable on
this same manifold which each exhibit compact Cauchy horizons of the
desired type. To stay on firm mathematical ground though we shall
only cite known results---realizing that these may be of lower dimension
than one might prefer.

\section{Kaluza-Klein Models for Spacetime Changing Generators }

The Misner metric~\cite{Misner} on $\mathbb{R}\times S^{1}$, expressed
in coordinates $\{t,\theta\}$, where $t\in\mathbb{R}$ and $\theta$
(defined mod $2\pi$) is a standard angle coordinate on the circle,
is given by
\begin{equation}
g_{M}=dt\otimes d\theta+d\theta\otimes dt-td\theta\otimes d\theta.\label{eq: g_M}
\end{equation}
or, in matrix form, 
\[
g_{M}=\left(\begin{array}{cc}
0 & 1\\
1 & -t
\end{array}\right).
\]
~\linebreak{}
Note that the Killing field $\frac{\partial}{\partial\theta}$ satisfies
$\frac{\partial}{\partial\theta}\cdot\frac{\partial}{\partial\theta}=(g_{M})_{\theta\theta}=-t$
and thus is spacelike for $t<0$, null at $t=0$ and timelike when
$t>0$. A closer inspection shows that the region $t<0$ of this Lorentzian
cylinder is globally hyperbolic, while the circle at $t=0$ serves
as its Cauchy horizon and the complementary region, $t>0$, is acausal,
with the orbits of $\frac{\partial}{\partial\theta}$ yielding closed
timelike curves. Misner's model may be viewed as a quotient of $2$-dimensional
Minkowski space by a Lorentzian boost (see pages 171--174 of reference~\cite{Hawking + Ellis - The large scale structure of spacetime}).

~

Taking the product of Misner's space with a flat, Riemannian $3$-torus
$\{T^{3},e\}$, where $e=\sum_{i=1}^{3}d\theta^{i}\otimes d\theta^{i}$
(with each $\theta^{i}$ a standard angle coordinate on the circle),
one arrives at the smooth, globally Lorentzian $5$-manifold $\{T^{3}\times\mathbb{R}\times S^{1},\tilde{g}\}$
with 
\begin{equation}
\tilde{g}=\sum_{i=1}^{3}d\theta^{i}\otimes d\theta^{i}+dt\otimes d\theta+d\theta\otimes dt-td\theta\otimes d\theta\label{eq: Lorentzian 5-metric}
\end{equation}
or, in matrix form (after relabeling via $\theta^{i}\longrightarrow x^{i}$,
$i=1,2,3$, $x^{4}=t$ and $x^{5}=\theta$)

\[
\tilde{g}=\left(\begin{array}{ccccc}
1 & 0 & 0 & 0 & 0\\
0 & 1 & 0 & 0 & 0\\
0 & 0 & 1 & 0 & 0\\
0 & 0 & 0 & 0 & 1\\
0 & 0 & 0 & 1 & -t
\end{array}\right).
\]
~\linebreak{}
This $5$-manifold is flat (and thus trivially satisfies the Einstein
equations) but only globally hyperbolic on the open submanifold $t<0$
while having a Cauchy horizon $\simeq T^{4}$ at $t=0$ and acausal
extension on the region $t>0$. Now, adopting the Kaluza-Klein parametrization
for the metric $\tilde{g}$ (in coordinates adapted to the Killing
field $\frac{\partial}{\partial\theta}=\frac{\partial}{\partial x^{5}}$)
one writes 

\begin{equation}
\tilde{g}=\left[\begin{array}{cc}
g_{\mu\nu}+\Phi A_{\mu}A_{\nu} & \Phi A_{\mu}\\
\Phi A_{\nu} & \Phi
\end{array}\right],\label{eq: KK matrix}
\end{equation}
wherein 

\begin{equation}
g={\displaystyle \sum_{\mu,\nu=1}^{4}}g_{\mu\nu}dx^{\mu}\otimes dx^{\nu}\label{eq: KK 4-metric}
\end{equation}
may (on the complement of the Cauchy horizon at $x^{4}=t=0$) be identified
with a (signature changing) metric on the $4$-dimensional quotient
manifold $\approx T^{3}\times\mathbb{R}$ and the one-form 
\begin{equation}
A={\displaystyle \sum_{\nu=1}^{4}}A_{\nu}dx^{\mu}\label{eq: vector potential}
\end{equation}
(in suitable electromagnetic units) with the vector potential of a
Maxwell field, while $\Phi$ is a scalar field on this same base.
Invariance of $\tilde{g}$ with respect to the $U(1)$ action generated
by $\frac{\partial}{\partial x^{5}}=\frac{\partial}{\partial\theta}$
ensures that these base fields $\{g,A,\Phi\}$ only depend on the
coordinates $\{x^{\mu};\mu=1,\ldots4\}$ of the base manifold.

~

In conventional Kaluza-Klein theory our $\Phi$ is often expressed
as $\Phi=\varphi^{2}$ since $\frac{\partial}{\partial\theta}$ is
there assumed to be uniformly spacelike and since the corresponding
scalar field $\varphi$ then satisfies a natural covariant wave equation
on the (uniformly Lorentzian) base manifold. For us though $\varphi$
would need to transition from real to imaginary to allow for the corresponding
transition of $\frac{\partial}{\partial\theta}=\frac{\partial}{\partial x^{5}}$
from spacelike to timelike whereas $\Phi$ need only change sign.
Also in conventional theory, the base fields $\{g,A,\varphi\}$ are
typically globally smooth (and satisfy a variant of the Einstein-Maxwell-scalar
field equations), but, in our setting, these fields will exhibit singularities
at the interface between Lorentzian and Riemannian geometry induced
upon the base $4$-manifold. To see this, note that, even for the
simple Misner model defined above, one has (with $x^{4}=t$ as above)

\begin{equation}
(g_{\mu\nu})=\left(\begin{array}{cccc}
1 & 0 & 0 & 0\\
0 & 1 & 0 & 0\\
0 & 0 & 1 & 0\\
0 & 0 & 0 & \frac{1}{t}
\end{array}\right),\label{eq: signature-change matrix}
\end{equation}

\begin{equation}
(A_{\nu})=(0,0,0,-\frac{1}{t}),\label{eq: singular vector potential}
\end{equation}

\begin{equation}
\Phi=-t=\varphi^{2}.\label{eq: singular scalar}
\end{equation}

\begin{rem}
The metric 
\[
g={\displaystyle \sum_{i=1}^{3}}dx^{i}\otimes dx^{i}+\frac{1}{t}dt\otimes dt
\]
exhibits a change of signature but is non-smooth, featuring an infinite
discontinuity at $t=0$. However, this discontinuity is merely a coordinate
singularity, as we will clarify below. This behavior contrasts with
the canonical smooth, transverse type-changing metrics considered
in the literature (see~\cite{Hasse + Rieger-Transformation, Kossowksi + Kriele - Signature type change and absolute time in general relativity, Kossowski - The Volume BlowUp and Characteristic Classes for Transverse, Rieger+Hasse-Loops}),
which take the form
\[
\bar{g}={\displaystyle \sum_{i=1}^{3}}dX^{i}\otimes dX^{i}+TdT\otimes dT,
\]
where the change of signature occurs smoothly along a hypersurface.
Our objective is to identify a coordinate transformation that maps
the non-smooth metric $g$ into the smooth representative $\bar{g}$.
Focusing on the time component of the metric, we impose the condition
$\frac{1}{t}(dt)^{2}=T(dT)^{2}.$ This yields the differential equation
$\left(\frac{dt}{dT}\right)^{2}=Tt$. Solving, we obtain a solution

\[
t(T)=\left(\pm\frac{1}{3}T^{\frac{3}{2}}\right)^{2}=\frac{1}{9}T^{3}.
\]
Substituting this back, we recover the smooth metric $\bar{g}$. In
this coordinate system, the corresponding fields are

\[
\Phi=-\frac{1}{9}T^{3}=-t=\varphi^{2},
\]
and
\[
A_{\nu}dx^{\nu}=-\frac{1}{t}dt=-\frac{3}{T}dT=A_{\nu'}dx^{\nu'}.
\]
These conditions ensure that, in the new coordinates, the metric $\bar{g}$
and the scalar field $\Phi$ exhibit a smooth, transverse type-changing
structure, while the vector field $A$ remains pure gauge.
\end{rem}

Thus, even though $\tilde{g}$ was smooth and Lorentzian throughout,
$g$ transitions from Lorentzian (for $t<0$) to Riemannian (with
$t>0$) across the interface at $t=0$.

~\linebreak{}
Note as well that whereas the base fields $\{g,A,\varphi\}$ will
(as a consequence of the imposed Ricci-flatness of $\tilde{g}$) automatically
satisfy the conventional (Kaluza-Klein) field equations on the Lorentzian
component of the base manifold, they will now transition to satisfying
Riemannian signature analogues of these equations on the acausal component
of the base whereon the equations become essentially elliptic instead
of hyperbolic. In the Misner model, for example, one easily checks
that $g$ (where defined) is flat, the electromagnetic Faraday tensor
$F=dA$ vanishes and $\varphi=\sqrt{-t}$ satisfies the wave equation,
$\oblong_{g}\varphi=0$, in the Lorentzian region $(t<0)$ but transitions
to a (purely imaginary) solution to Laplace's equation, $\triangle_{g}\varphi=0$,
on the acausal extension $(t>0)$.

~

Lest the reader assume that such (type-changing, Kaluza-Klein) solutions are limited to small families of the Misner variety we sketch in the section below a technique for constructing infinite dimensional sets of such type-changing solutions.

~\linebreak{}
By appealing, for example, to the Cauchy-Kowalewski theorem one expects
to create much larger families of higher dimensional, Einstein spacetimes
exhibiting the same (``signature change without signature change'')
behavior but, so far as are know, this has only, until now, been carried
out explicitly in the (lower dimensional) context of $4$-dimensional,
$U(1)$-symmetric, Einstein spacetimes over (signature changing)
$3$-dimensional quotients\LyXZeroWidthSpace ~\cite{6 Moncrief ,5 The space of generalized Taub-Nut spacetimes}.

~\linebreak{}
For this reason let us fall back by $1$ dimension and, as in Ref.~\cite{6 Moncrief },
consider Lorentzian metrics defined on manifolds of the form $^{(4)}V=K\times\mathbb{R}\times S^{1}$,
where $K$ is a compact, connected, orientable surface. One views
these as (trivial) circle bundles over the base manifolds $K\times\mathbb{R}$
and imposes upon the metrics to be considered the isometry group of
$U(1)$-invariance under translations along the circular fibers.
For simplicity, we shall only treat trivial (i.e., product) bundles
here, but the same techniques are applicable to nontrivial $S^{1}$-bundles
such as $S^{3}\times\mathbb{R}\longrightarrow S^{2}\times\mathbb{R}$
as discussed in Ref.~\cite{5 The space of generalized Taub-Nut spacetimes}.

~

Let $\{x^{a},a=1,2\}$ represent local coordinates on $K$, $x^{3}=\theta$
(defined mod $2\pi$) be an angle coordinate on the circle and $x^{0}=t$
designate the ``time''. Consider analytic, Lorentzian metrics on
$^{(4)}V$ expressible as 
\begin{equation}
\tilde{g}=e^{-2\lambda}\{\frac{(N^{2}-e^{4\lambda})}{t}dt\otimes dt+\sum_{a,b=1}^{2}{}^{(2)}g_{ab}dx^{a}\otimes dx^{b}\}-te^{2\lambda}(d\theta+\alpha_{a}dx^{a})\otimes(d\theta+\alpha_{b}dx^{b})\label{eq: analytic Lorentzian metric}
\end{equation}

\[
+e^{2\lambda}\{dt\otimes(d\theta+\alpha_{a}dx^{a})+(d\theta+\alpha_{a}dx^{a})\otimes dt\},
\]
where $\frac{\partial}{\partial\theta}$ is a Killing field so that
the various metric components depend only upon $\{t,x^{1},x^{2}\}$.\footnote{Where the coordinates employed here are (constant multiples of) the
primed coordinates $\{t^{\prime}, x^{3\prime}, x^{a\prime}\}$ used
previously in~\cite{6 Moncrief }.} In the above 
\begin{equation}
^{(2)}g={\displaystyle \sum_{a,b=1}^{2}}{}^{(2)}g_{ab}dx^{a}\otimes dx^{b}\label{eq: analystic 2-metric}
\end{equation}
is (at each fixed $t$) a Riemannian metric expressed in local charts
for $K$ and we have (without any essential loss of generality) taken
the shift field to vanish so that $\tilde{g}$ is parametrized by
only $7$ (instead of the usual $10$) functions $\{N,\lambda,^{(2)}g_{ab},\alpha_{a}\}$.
The above metric will be analytic and Lorentzian on at least a neighborhood
$\mathcal{N}=K\times S^{1}\times(-\rho,\rho)$ of the hypersurface
$t=0$ provided 

~
\begin{lyxlist}{00.00.0000}
\item [{(i)}] $\{\lambda,N,\alpha_{a},^{(2)}g_{ab}$ are analytic on $\mathcal{N}$,\\
\item [{(ii)}] $N>0$ and $^{(2)}g$ is Riemannian on $\mathcal{N}$, and\\
\item [{(iii)}] $\left(\frac{N^{2}-e^{4\lambda}}{t}\right)$ is analytic
on $\mathcal{N}$. 
\end{lyxlist}
~

By examining the metric in more detail, one can verify that\\

\begin{lyxlist}{00.00.0000}
\item [{(iv)}] the hypersurface $t=0$ is a null hypersurface with the
Killing field $\frac{\partial}{\partial\theta}$ tangent to its null
generators, and\\
\item [{(v)}] the Killing field $\frac{\partial}{\partial\theta}$ is spacelike
in the region $t<0$ but timelike in the complementary region $t>0$
where its orbits are closed timelike curves.
\end{lyxlist}
~\linebreak{}
Spacetimes satisfying conditions (i)--(iii) above are globally hyperbolic
in the regions $t<0$, have Cauchy horizons diffeomorphic to $K\times S^{1}$
at $t=0$ and are acausal in the regions $t>0$.

~

If, as in Refs.~\cite{6 Moncrief ,5 The space of generalized Taub-Nut spacetimes},
we impose Einstein's vacuum field equations upon metrics of the form
(\ref{eq: analytic Lorentzian metric}) then we may prove the existence
of infinite-dimensional families of solutions having all the properties
(i)--(v) above provided we impose a suitable coordinate condition
to fix the lapse function $N$ (recalling that the shift field has
already been set to vanish). The basic steps in the proof are an application
of the generalized Cauchy-Kowalewski theorem sketched in Ref.~\cite{6 Moncrief }
and proven in detail in Ref.~\cite{5 The space of generalized Taub-Nut spacetimes}.\footnote{The need for an extension of the classical Cauchy-Kowalewski theorem
arises because of the occurrence of so-called Fuchsian singularities
in the field equations for metrics of the type under consideration.}

~

The main result is that every choice of analytic initial data $\{\mathring{\lambda},\mathring{\alpha}_{a},^{(2)}\mathring{g}_{ab}\}$
$(0,x^{1},x^{2})$ specified over $K$ (with $\mathring{\varphi}$
a function, $\mathring{\alpha}_{a}dx^{a}$ a one-form and $^{(2)}\mathring{g}_{ab}dx^{a}x^{b}dx^{b}$
a Riemannian metric) determines a unique, analytic solution of the
vacuum Einstein equations having all the properties (i)--(v) above,
provided that the lapse function is chosen to satisfy conditions (i)--(iii)
above and the additional coordinate condition 

~
\begin{lyxlist}{00.00.0000}
\item [{(vi)}] $\left(\frac{N}{\sqrt{\det^{(2)}g}}\right)_{,t}=0$\label{6 additional coordinate condition }, 
\end{lyxlist}
~\linebreak{}
where $\det^{(2)}g$ is the determinant of $^{(2)}g$. Together these
restrictions lead to the requirement that

\begin{equation}
\frac{N}{\sqrt{\det^{(2)}g}}=\frac{e^{2\lambda}}{\sqrt{\det^{(2)}\mathring{g}}},\label{eq: requirement}
\end{equation}
which fixes $N$ completely in terms of the remaining variables.

~\linebreak{}
These rigid coordinate conditions {[}i.e., zero shift together with
(\ref{eq: requirement}){]} are not strictly necessary but were chosen
to simplify the form of Einstein's equations and to facilitate the
application of the generalized Cauchy-Kowalewski theorem in Refs.~\cite{6 Moncrief }
and~\cite{5 The space of generalized Taub-Nut spacetimes}.

~

Many of the solutions determined by data $\{\mathring{\lambda},\mathring{\alpha}_{a},^{(2)}\mathring{g}_{ab}\}$
prescribed on $K$ will be isometric to one another. For any such
solution however, one can, without disturbing the coordinate conditions
imposed above, find a diffeomorphism of $^{(4)}V$ that takes $\tilde{g}$
to a canonical gauge in which

~
\begin{lyxlist}{00.00.0000}
\item [{(a)}] \noindent $^{(2)}\mathring{g}_{ab}dx^{a}\otimes dx^{b}$
is a constant curvature metric on $K$ depending only on the choice
of zero (if $K\approx S^{2})$, two (if $K\approx T^{2})$, or $6g-6$
(if $K$ has genus $g\geq2$) real parameters;\\
\item [{(b)}] \noindent $\alpha_{a}dx^{a}$ has vanishing divergence with
respect to $^{(2)}\mathring{g}_{ab}dx^{a}\otimes dx^{b}$; and\\
\item [{(c)}] \noindent there is a residual gauge subgroup action of dimension
$6$ (if $K\approx S^{2}$) or dimension $2$ (if $K\approx T^{2})$
generated by the conformal Killing fields of $\{K,^{(2)}\mathring{g})$
that act on the data $\{\mathring{\lambda},\mathring{\alpha}_{a}dx^{a},^{(2)}\mathring{g}_{ab}dx^{a}\otimes dx^{b}\}$.
Thus $\mathring{\lambda}$ and the divergence-free component of $\mathring{\alpha}_{a}dx^{a}$
together with the Teichm\"uller parameters for $^{(2)}\mathring{g}$
(modulo the action of a finite dimensional Lie group in the case $K\approx S^{2}$
or $T^{2}$) represent the truly independent data that parametrize
the nonisometric solutions of Einstein's equations on $^{(4)}V$ which
admit compact Cauchy horizons of the type described above.
\end{lyxlist}
~

Given any such solution though one can now derive the fields $\{g,A,\Phi\}$
induced upon the base manifold, $^{(4)}V/U(1)\approx K\times\mathbb{R}$,
through an application of the $4$-dimensional version of formula
(\ref{eq: KK matrix}). The result is easily found to be:

\begin{equation}
\Phi=-te^{2\lambda}=\varphi^{2},\label{eq: new scalar}
\end{equation}

\begin{equation}
A=-\frac{1}{t}dt+\alpha_{a}dx^{a},\label{eq: new vector potential}
\end{equation}

\begin{equation}
g=g_{\mu\nu}dx^{\mu}\otimes dx^{\nu}=e^{-2\lambda}\left\{ \frac{N^{2}}{t}dt\otimes dt+^{(2)}g_{ab}dx^{a}\otimes dx^{b}\right\} \label{eq: new metric}
\end{equation}

~\\
wherein, of course, we are now applying the Einstein summation convention
to simplify the notation. Notice that $\varphi$, $A$ and $g$ are
each singular at the interface $t=0$ at which $g$ transitions from
being Lorentzian (for $t<0$) to Riemannian (for $t>0$).

~

While the foregoing examples, aside from the $5$-dimensional Misner
model, deal only with $4$-dimensional, trivial circle bundles over
$3$-dimensional (signature changing) quotients, there is good reason
to suppose that the analysis can be extended to cover nontrivial bundles
and higher-dimensional bundles over a variety of bases. Indeed, the
case of $S^{3}\times\mathbb{R}\rightarrow S^{2}\times\mathbb{R}$
(involving the Hopf fibration of $S^{3}$ over $S^{2}$) has already
been treated and yields an infinite dimensional extension of the $2$-parameter
family of classical Taub-NUT solutions~\cite{5 The space of generalized Taub-Nut spacetimes}.

~

A key point is that the generalized Cauchy-Kowalewski theorem is insensitive
to dimension and the Fuchsian singularities in the higher dimensional
Einstein field equations are expected to have the same form as those
we have already treated in $4$-dimensions. On the other hand, imposing
field equations at all is a much more constrained arena for studying
signature change than the more abstract approach would usually consider,
so the reader may well wonder whether any advantages accrue from this
more circumscribed scenario.

~

To address this question, recall that whereas in the abstract approach
there is no difficulty in defining geodesic curves in the purely Lorentzian
or purely Riemannian components of a signature-changing manifold $M$,
the continuation of such curves across a hypersurface $\mathcal{H}\subset M$
of signature change can be problematic since the Levi-Civita connection
components, entering crucially in the geodesic equations, fail to
be defined on $\mathcal{H}$.

~

In the Kaluza-Klein framework, though, there is no difficulty in defining
geodesics up in the bundle where the metric ($\hat{g}$ in our notation)
is globally smooth and Lorentzian. But what do such geodesic curves
``upstairs'' have to do with geodesics down in the quotient space,
even in those regions where such a notion is well-defined? Though
the answer is well-known within conventional Kaluza-Klein theory,
it is worth recalling here.

~

The Lagrangian for (say) the timelike geodesics of a massive particle
living up in the bundle is given by (see also~\cite{Kerner et al})

\begin{equation}
L=\frac{1}{2}m\hat{g}_{\eta\nu}\frac{dx^{\mu}}{d\lambda}\frac{dx^{\nu}}{d\lambda},\label{eq: Lagrangian}
\end{equation}
where $m>0$ is the (constant) mass, and $\lambda$ is an affine parameter
along the curve (e.g., proper time). Reexpressing this, though, in
terms of the base fields via equation (\ref{eq: KK matrix}) yields

\begin{equation}
L=\frac{1}{2}m\left\{ g_{\mu\nu}\frac{dx^{\mu}}{d\lambda}\frac{dx^{\nu}}{d\lambda}+\Phi\left(\frac{dx^{5}}{d\lambda}+A_{\mu}\frac{dx^{\mu}}{d\lambda}\right)^{2}\right\} ,\label{eq: explicit Lagrangian}
\end{equation}
which clearly leads to non-geodesic motion relative to the base metric
$g_{\mu\nu}dx^{\mu}\otimes dx^{\nu}$ unless the contribution of $\Phi$
and $A_{\mu}dx^{\mu}$ to be equations of motion can be suppressed.\footnote{A related analysis of geodesic deviation and motion in Kaluza-Klein theories was carried out in Ref.~\cite{Kerner et al} where the influence of the higher-dimensional geometry on the effective 4-dimensional dynamics was studied in detail.}
There is, however, a well-known way of doing this. Since $x^{5}$
is a cyclic coordinate (due to the fact that $\frac{\partial}{\partial x^{5}}$
is Killing), its conjugate momentum, 
\begin{equation}
p_{5}=\frac{\partial L}{\partial(\frac{dx^{5}}{d\lambda})}=m\Phi\left(\frac{dx^{5}}{d\lambda}+A_{\mu}\frac{dx^{\mu}}{d\lambda}\right),\label{eq: cyclic coordinate}
\end{equation}
is a constant of the motion ($\frac{dp_{5}}{d\lambda}=\frac{\partial L}{\partial x^{5}}=0$)
which plays the role of electric charge in the associated Lorentz
force equation. Setting this constant to zero reduces the Euler-Lagrange
equations to geodesic form, at least in those regions of the quotient
manifold wherein the base fields $\{\Phi,A,g\}$ are well-defined.
But now these curves, viewed as geodesics of the globally smooth metric
$\hat{g}$, have no difficulty crossing the interface, which, upstairs,
is nothing but the Cauchy horizon of an Einstein spacetime.

~

On the other hand though, the interpretation of these (Euler-Lagrange)
solution curves as geodesics in the bundle does not, in general, descend
to apply to the corresponding curves down in the quotient, base manifold,
especially when these curves in the bundle cross the Cauchy horizon.
To satisfy the pure geodesics equations in the base the solution curves
up in the bundle must, as we have mentioned, have a vanishing value
of $p_{5}$, the constant of motion which plays the role of electric
charge.

~

But, for the spacetimes under consideration here\LyXZeroWidthSpace{}
(c.f., Eqs.(\ref{eq: cyclic coordinate}, \ref{eq: new scalar}, \ref{eq: new vector potential},
\ref{eq: new metric})) the vanishing of 

\[
p_{5}=m\Phi\left(\frac{dx^{5}}{d\lambda}+A_{\mu}\frac{dx^{\mu}}{d\lambda}\right)=-me^{\lambda}\left(t\frac{dx^{5}}{d\lambda}-\frac{dt}{d\lambda}+t\alpha_{a}\frac{dx^{a}}{d\lambda}\right)
\]
for a solution curve that crosses the horizon at $t(\lambda_{*})=0$
transversally, with $\frac{dt}{d\lambda}\mid_{\lambda=\lambda_{*}}\neq0$,
would have to have 

\[
\frac{dx^{5}}{d\lambda}\underset{\lambda\longrightarrow\lambda_{*}}{\longrightarrow}\pm\infty
\]
and thus not actually extend to the horizon after all. Thus, unfortunately,
our realization of signature changing manifolds via an extension of
the Kaluza-Klein paradigm does not help to resolve the question of
how, naturally, to extend geodesics across a singular hypersurface
in the base.

~

A somewhat related question is whether the quotient manifold with
metric $g$ (in regions where this is well-defined) can be realized
as an isometric embedding of a cross section of the $S^{1}$-bundle,
with metric $\tilde{g}$. The well-known answer is that this is not
the case unless the curvature of this bundle, represented by the Faraday
tensor $F_{\mu\nu}dx^{\mu}\wedge dx^{\nu}$, vanishes. To see this,
note that in the chosen coordinates, a cross section would be defined
by setting $x^{5}=\Lambda(t,x^{1},\ldots,x^{3})$ for some smooth
function $\Lambda$. But the metric induced by $\tilde{g}$ upon this
cross section would agree with $g$ (where the latter is defined)
if and only if $A_{\mu}+\frac{\partial\Lambda}{\partial x^{\mu}}=0$,
i.e., if and only if the vector potential is pure gauge and thus its
corresponding exterior derivative,$F$, vanishes.

~

Finally, the type-changing (Einstein-Maxwell scalar) geometric field
equations down in the quotient manifold could be problematic to analyze
directly (especially when their solutions are expected to be singular
at the signature-changing interface), but in this extended Kaluza-Klein
setting, they are singular projections of a globally smooth metric
up in the bundle where nothing really singular actually happens (except
the loss of global hyperbolicity upon crossing the horizon).

\section{Cauchy Horizons as Killing Horizons }

The constructions described in the previous section of $U(1)$-symmetric,
analytic, vacuum spacetimes having compact Cauchy horizons led to
the suspicion early on that the presence of the $U(1)$-generating
Killing field (which was tangent to the horizon's null generators)
was actually necessary for the Cauchy horizon's existence rather
than being merely a simplifying ansatz to make for an interesting
special case. That analytic, compact Cauchy horizons were necessarily
Killing horizons (for solutions to the electrovacuum field equations)
was then proven by J. Isenberg and one of us in the special case that
the horizon's null generators were all closed curves (that, in addition,
satisfied a local product bundle condition)~\cite{1 Symmetries of cosmological Cauchy horizons}.\footnote{This condition excluded for example, Seifert fibered horizons admitting
exceptional fibers around which the generic fiber spins in barberpole
fashion. Eventually though this hypothesis was eliminated~\cite{4 On spacetimes containing Killing vector fields with non-closed orbits,2 Symmetries of Cosmological Cauchy Horizons with Non-Closed Orbits}.} While the assumption of closure of the null generators seemed at
first to be an artificially restrictive condition to impose on a Killing
horizon, these same authors later showed that non-closure of the generators
implied the presence of an independent Killing field that commuted
with the  horizon generating one so that, together, they generated
a full $T^{2}$ isometry group action on the enveloping Einstein spacetime.
Examples were also known, which these same authors referred to as
``ergodic'', in which the null generators densely filled the entire
($3$-dimensional) Cauchy horizon and a corresponding, highly rigidifying,
$T^{3}$ isometry group of the spacetime was then in play.\footnote{By making irrational shifts in the identifications that toroidally
compactify the ``flat Kasner'' Einstein spacetime, one can easily
construct examples in which the Cauchy horizon $\approx T^{3}$ is
densely filled by each of its null generators. These were conjectured
to exhaust the ergodic cases, a result that was later proven by Bianchi
and Reisis~\cite{A classification theorem for compact Cauchy horizons in vacuum spacetimes}.}

Finally, setting aside the ergodic cases, these same authors showed
that the generic non-closed generators for analytic, vacuum Cauchy
horizons densely filled $2$-tori embedded in the horizon~\cite{4 On spacetimes containing Killing vector fields with non-closed orbits,2 Symmetries of Cosmological Cauchy Horizons with Non-Closed Orbits}.
More recently, a number of researchers have significantly extended
the known results on compact Cauchy horizons in vacuum or non-vacuum
spacetimes. Since most of these results, though, lie outside the Kaluza-Klein
paradigm of a higher-dimensional spacetime admitting a $U(1)$-generating
Killing field, we shall not attempt to review them here.

~

One development worth recalling in this context, though, is that ``cosmological''
spacetimes admitting compact Cauchy horizons can often be created
by taking suitable quotients of stationary black holes in $4$ and
higher dimensions. The symmetry groups of higher-dimensional black
objects (e.g., black holes, black rings, black Saturns, etc.) and
their connections to the closure or non-closure of the generators
of these objects' compactified horizons have been analyzed in detail
elsewhere~\cite{Symmetries of higher dimensional black holes}.

~
\begin{acknowledgement*}
Moncrief is especially grateful to the Erwin-Schr\"{o}dinger Institute
of the University of Vienna for hospitality and support where part
of this research was carried out. Rieger gratefully acknowledges support
from the Simons Center for Geometry and Physics, Stony Brook University,
where part of this work was conducted.
\end{acknowledgement*}
~

\end{document}